\documentclass[a4paper,12pt]{article}
\usepackage{jheppub,esint,psfrag}
\usepackage[utf8]{inputenc}

\usepackage{epsfig,amssymb,amsmath,psfrag,subfigure,rotate,color,wasysym,xcolor}
\usepackage[bbgreekl]{mathbbol}

\allowdisplaybreaks 
\newcommand{\insertfig}[2]{\includegraphics[width=#1cm]{#2}}

\DeclareSymbolFontAlphabet{\mathbbm}{bbold}
\DeclareSymbolFontAlphabet{\mathbb}{AMSb}%

\def\XXint#1#2#3{{\setbox0=\hbox{$#1{#2#3}{\int}$ }
\vcenter{\hbox{$#2#3$ }}\kern-.6\wd0}}

\def \be  {\begin{equation}}
\def \ee  {\end{equation}}
\def \ba  {\begin{eqnarray}}
\def \ea  {\end{eqnarray}}
\def \baa {\begin{eqnarray*}}
\def \eaa {\end{eqnarray*}}
\newcommand{\ep}{\varepsilon}
\def \lab #1 {\label{#1}}

\newcommand\re[1]{(\ref{#1})}
\def\d{\hbox{{d}\kern-.20em\hbox{l}}}

\def \matrix #1 {\left(\begin{array}{cc} #1 \end{array}\right)}

\def \tr {\mathop{\rm tr}\nolimits}

\def\1{\hbox{{1}\kern-.25em\hbox{l}}}

\newcommand{\ft}[2]{{\textstyle\frac{#1}{#2}}}

\makeatletter

\input pdf-trans
\newbox\qbox
\def\usecolor#1{\csname\string\color@#1\endcsname\space}
\newcommand\bordercolor[1]{\colsplit{1}{#1}}
\newcommand\fillcolor[1]{\colsplit{0}{#1}}
\newcommand\outline[1]{\leavevmode%
  \def\maltext{#1}%
  \setbox\qbox=\hbox{\maltext}%
  \boxgs{Q q 2 Tr \thickness\space w \fillcol\space \bordercol\space}{}%
  \copy\qbox%
}
\makeatother
\newcommand\colsplit[2]{\colorlet{tmpcolor}{#2}\edef\tmp{\usecolor{tmpcolor}}%
  \def\tmpB{}\expandafter\colsplithelp\tmp\relax%
  \ifnum0=#1\relax\edef\fillcol{\tmpB}\else\edef\bordercol{\tmpC}\fi}
\def\colsplithelp#1#2 #3\relax{%
  \edef\tmpB{\tmpB#1#2 }%
  \ifnum `#1>`9\relax\def\tmpC{#3}\else\colsplithelp#3\relax\fi
}
\bordercolor{black}
\fillcolor{white}
\def\thickness{.3}

\def\1{\mathbbm{1}}

\title{Off-shell minimal form factors}
 
\author[a]{A.V.~Belitsky,}
\author[b]{L.V.~Bork}
\affiliation[a] {Department of Physics, Arizona State University,  Tempe, AZ 85287-1504, USA}  
\affiliation[b]{Institute for Theoretical and Experimental Physics, 117218 Moscow, Russia}
\affiliation[]{The Center for Fundamental and Applied Research, 127030 Moscow, Russia}
  
 \abstract
{We study off-shell $n$-particle form factors of half-BPS operators built from $n$ complex scalar fields at the two-loop order
in the planar maximally supersymmetric Yang-Mills theory (sYM). These are known as minimal form factors. We construct their
representation as a sum of independent scalar Feynman integrals relying on two complementary techniques. First, by going 
to the Coulomb branch of the theory by employing the spontaneous symmetry breaking which induces masses, but only for 
external particles while retaining masslessness for virtual states propagating in quantum loops. For a low number of external
legs, this entails an uplift of massless integrands to their massive counterparts. Second, utilizing the $\mathcal{N}=1$ superspace 
formulation of $\mathcal{N} = 4$ sYM and performing algebra of covariant derivatives off-shell. Both techniques provide
identical results. These form factors are then studied in the near-mass-shell limit with the off-shellness regularizing emerging
infrared divergences. We observe their exponentiation and confirm the octagon anomalous dimension, not the cusp, as the 
coefficient of the Sudakov double logarithmic behavior. By subtracting these singularities and defining a finite remainder, we 
verified that its symbol is identical to the one found a decade ago in the conformal case. Beyond-the-symbol contributions are  
different in the two cases, however.}
 
\begin{document}

\maketitle
\flushbottom
\setcounter{footnote} 0

\section{Introduction}
\label{s1}

Multiparticle form factors are intrinsic building blocks in the construction of physical observables measurable in a variety of 
high-energy processes. They are defined by an overlap of a state sourced by a gauge-invariant composite operator 
$\mathcal{O} (x)$ from the vacuum and a prepared $n$-particle state, namely,
\begin{align}
\label{MatrixElement}
\int d^4 x \, {\rm e}^{- i q \cdot x} \langle p_1, {\dots} , p_n | \mathcal{O} (x) | 0 \rangle
=
(2 \pi)^4 \delta^{(4)} (p_1 + {\dots} + p_n - q) \mathcal{F}_n
\, .
\end{align}
Massless gauge theories are plagued by uncontrolled emission of soft gauge bosons, leading to the infrared catastrophe. Since 
the final state in the form factor is fixed, $\mathcal{F}_n$ takes on the role of a probability of no additional emission from the source. 
This implies that the form factors per se are not physical quantities\footnote{Nevertheless, we will call them `observables'.} and suffer 
from the aforementioned divergencies. Only in the sum with additional contributions involving real emissions of gauge bosons do the 
latter cancel yielding a finite remainder that can be confronted with experimental data. This is the basis of the Kinoshita-Lee-Nauenberg 
theorem \cite{Kinoshita:1962ur,Lee:1964is}. 

QCD factorization theorems \cite{Collins:1989gx} are at the core of the systematic separation of different momentum scales involved 
in a physical process and their calculability in terms of the parton-level scattering. Depending on the kinematical situation, however, 
partons can be either on or off their mass shell. For instance, collinear factorization \cite{Collins:1989gx} implies the former, while the 
high-energy \cite{Catani:1990eg,Collins:1991ty} one does the latter. Naively taking partonic scattering off the mass shell breaks its gauge 
invariance, thus, hampering straightforward uses of the Feynman diagram technique. A way out of this predicament is to split up 
fields in terms of `classical' and `quantum', with the former appearing in external states and the latter in virtual loops only. By imposing 
physical gauges (like Fock-Schwinger or light-cone gauges) on external potentials, the internal fields are then constrained with 
background covariant gauges \cite{Abbott:1980hw,Abbott:1981ke}. The two gauges are unrelated and allow one to express all external 
gauge potentials in terms of the field strength tensors and obtain gauge invariant off-shell scattering amplitudes \cite{Gates:1983nr}.

The Feynman diagram technique is, however, highly impractical for loop calculations and was largely superseded by unitarity-cut 
techniques \cite{Bern:1994zx,Bern:1994cg,Britto:2004nc,Cachazo:2008vp} and spinor-helicity formalisms \cite{Gastmans:1990xh,Xu:1986xb} 
in recent decades. The first of these allows one to operate only in terms of on-shell amplitudes, while the latter provides them in a highly 
compact form \cite{Parke:1986gb} making this a go-to tool in modern-day applications. These frameworks were instrumental in the 
recent tremendous computational progress of higher order corrections in QCD and its supersymmetric siblings like $\mathcal{N} = 4$ 
supersymmetric Yang-Mills (sYM), see, e.g., \cite{Dixon:2013uaa}. 

But what about an off-shell generalization of the above technique? Dimensional reconstruction appears to be the one to fill the void
\cite{Giele:2008ve,Ellis:2008ir,Bern:2010qa,Boughezal:2011br,Davies:2011vt,AccettulliHuber:2019abj}. This framework was used in 
the past few years to fix $D$-dimensional integrands of loop amplitudes with external massless legs bypassing the use of 
$D$-dimensional cuts which lose the benefits of the spinor-helicity representation for the building blocks and complicate intermediate 
calculations. Instead, one calculates the cuts in a few integer dimensions and then interpolates these to get integrands for generic $D$. 
Higher-dimensional generalizations of the spinor-helicity formalism were used in these applications. In particular, the one for six space-time 
dimensions \cite{Cheung:2009dc}. This approach allows one to calculate not only cut-constructible contributions but also the so-called
rational terms, invisible in four dimensions.

Recently, the above logic of dimensional reconstruction was inverted within the context of $\mathcal{N} = 4$ sYM \cite{Caron-Huot:2021usw}. 
Namely, to address the off-shell situation, one considers higher integer-dimensional generalization of particle scattering with null, i.e., 
massless, momenta. However, one interprets the out-of-four-dimensional components of these momenta as virtualities (or masses) 
making them off-shell (or massive) from the four-dimensional standpoint \cite{Selivanov:1999ie,Boels:2010mj}. This is where this procedure 
deviates from the conventional dimensional reduction which yields the conformal branch of the theory \cite{Brink:1976bc,Gliozzi:1976qd}. This 
setup possesses all of the advantages of the original four-dimensional one. The most important one is that it preserves the gauge 
invariance of higher-dimensional integrands. These can then be dimensionally reduced down to four. Since external state virtualities make 
loop integrals infrared finite, one can safely set out-of-four-dimensional components of loop momenta to zero in theories without ultraviolet 
divergencies and simultaneously set their dimension equal to four. This construction was used in the past couple of years to predict off-shell 
amplitudes \cite{Caron-Huot:2021usw,Bork:2022vat} and form factors \cite{Belitsky:2022itf,Belitsky:2023ssv,Belitsky:2024agy} in the 
$\mathcal{N} =4$ theory. For a relatively small number of external legs, these arguments allowed one to bypass explicit higher-dimensional 
analyses, since corresponding integrands coincide in four and $D$-dimensions, provided one assumes that all four-dimensional Lorentz 
products are regarded as $D$-dimensional ones. This is the so-called {\sl uplift} \cite{Huang:2011um,Plefka:2014fta}. It is important to 
realize that it is destined to fail since starting from a certain number of external legs, integrands begin to develop $(D-4)$-dimensional 
contributions, the so-called $\mu$-terms, see e.g., \cite{Bern:2008ap}, that can not be unambiguously uplifted.

Although the higher-dimensional perspective is undoubtedly the most effective way to access off-shell `observables', these can
nevertheless be addressed solely within the four-dimensional setup of the $\mathcal{N} = 4$ sYM. Now, masses will be 
generated by the conventional Higgs mechanism for some of the model's scalars \cite{HennGiggs1}. Since all of the scalars emerge 
from the dimensional reduction of a ten-dimensional adjoint gauge potential, thus induced vacuum expectation values, i.e., moduli, take 
the theory away from the origin of its moduli space and into its Coulomb branch. A proper choice of complex moduli can yield
only massive fields in external states and massless in internal propagators. The two viewpoints are equivalent.

One question that begs an answer is whether the limit when all moduli tend to zero is smooth. While one would 
naively assume this, it turns out to be far from reality. Nonvanishing moduli provide infrared regularization of amplitudes
and form factors. So, as one sends them to zero, the latter diverge and are not defined. To properly take this limit, one must 
take momentum integrals away from four dimensions. Having two regulators at the same time will the limits commute? This 
question received a negative answer in Ref.\ \cite{Caron-Huot:2021usw}  where four-gluon scattering amplitudes were  
analyzed making use of the uplift from $D$-dimensional cuts \cite{Bourjaily:2016evz}, it was shown that in the near-mass-shell 
limit, the infrared behavior of a four-gluon amplitude is driven by the so-called octagon anomalous dimension rather than 
the expected ubiquitous cusp. 

The above observation triggered a series of papers where we initiated systematic studies of `observables' on the Coulomb
branch of the $\mathcal{N} = 4$ sYM \cite{Belitsky:2022itf,Belitsky:2023ssv,Belitsky:2024agy}. Form factors were chosen as 
a case study, rather than amplitudes for technical reasons. Namely,  the former allows one to induce nontrivial finite remainder 
functions for a fewer number of legs as compared to amplitudes. For instance, a three-leg form factor already possesses one 
starting from the two-loop order, while for amplitudes one would have to consider at least six legs. Two- and three-leg form factors 
of operators from the stress-tensor multiplet were studied at three and two loops, respectively. Both considerations confirmed 
the octagon anomalous dimension as an off-shell surrogate of the on-shell cusp. Currently being at the `data' collection stage, 
we perform a two-loop calculation of the so-called off-shell minimal form factors. These are form factors of half-BPS
operators $\mathcal{O}$ with the same number of fields building them up as the number of external legs.

Our analysis is structured as follows. In the next section, we provide definitions for our `observables' and their anticipated
infrared properties. Next, we use uplifted four-dimensional integrands for one- and two-loop form factors. To confirm the 
emerging set of independent scalar Feynman integrals, we perform another derivation relying on the superspace formulation of 
the model. In the near mass-shell limit, we find that the two expressions coincide. In section \ref{ResultsSection}, we provide
expressions for all contributing integrals in terms of multiple polylogarithms. Some of them are extremely long and lack
uniform transcendentality. However, we demonstrate that they nicely combine into very compact expressions. Finally,
we conclude.

\section{Minimal form factors}

The main objects for our current analysis are the form factors \re{MatrixElement} generated by the half-BPS operators
\begin{align}
\label{phi12n}
\mathcal{O} = \tr \phi_{12}^n
\, ,
\end{align}
built from $n$ sextets of scalars $\phi_{AB}$ of the $\mathcal{N} = 4$ sYM. Here and below, we use matrix-valued fields with 
SU($N_c$) generators normalized as $\tr t^a t^b = \ft12 \delta^{ab}$. These `observables' were dubbed minimal form
factors in Ref.\ \cite{Brandhuber:2014ica} due to their property that they cannot have fewer external legs than the fields 
in $\mathcal{O}$. The latter are generalizations of the lowest component of the chiral part of the stress tensor 
supermultiplet $\mathcal{T}$ \cite{Eden:2011yp}
\begin{align}
\mathcal{T} (x, \theta_+) = \tr \phi_{++}^2 (x) + \dots + \theta_+^4 \mathcal{L} (x)
\, ,
\end{align}
where $\mathcal{L}$ is the chiral on-shell Lagrangian and, without loss of generality, one can choose harmonic projections of the SU(4) 
indices to be $\phi_{++} \to \phi_{12}$.

The form factors $\mathcal{F}_n$ describe the process of `creating' $n$ scalars of momenta $p_i$ by the source $\mathcal{O}$, which
has the off-shell Euclidean momentum ($q^2 < 0$)
\begin{align}
q = p_1 + {\dots} + p_{n}
\, .
\end{align}
They are functions of the Lorentz invariants $s_{ij}$ and the off-shellness $m$, which we choose the same for all external legs,
\begin{align}
\label{KinInvariants}
s_{ij} = (p_i + p_j)^2
\, , \qquad
m = - p_i^2
\, .
\end{align}
All invariants are taken in the Euclidean domain, i.e., $s_{ij} <0$ and $m >0$, to avoid dealing with imaginary parts due to particle 
production thresholds. We already addressed the two-particle case earlier up to the three-loop order \cite{Belitsky:2022itf,Belitsky:2023ssv}. 
Presently, we will perform the calculation of $\mathcal{F}_n$ at two loops,
\begin{align}
\mathcal{F}_n = 1 + g^2 \mathcal{F}^{(1)}_n + g^4 \mathcal{F}^{(2)}_n + O(g^6)
\, ,
\end{align}
where the planar perturbative expansion runs in 't Hooft coupling $g^2 = g^2_{\scriptscriptstyle\rm YM} N_c/(4 \pi)^2$.

Before we turn to the actual calculations, let us recall the general structure of form factors on the Coulomb branch and its conformal
limit. In the latter case, infrared singularities are regularized using dimensional reduction. These then emerge as poles in 
$\varepsilon = (D - 4)/2$. Since infrared physics can be reformulated in terms of expectation values of Wilson lines meeting at 
the point $x$ of the source $\mathcal{O} (x)$ insertion, the infrared poles get replaced by the ultraviolet ones. These come from
short-distance gluon exchanges in the vicinity of $x$, where Wilson lines form a cusp. These depend on the ratios of Lorentz invariants
and a mass scale $\mu$ of the dimensional regularization of loop integrals $\mu^2/s_{ij}$, however, planarity eliminates all contributions
for the exception of $j = i + 1$. As a consequence of this, the form factors admit the form
\begin{align}
\label{AmplFactor}
\log \mathcal{F}_{n} = \frac{1}{2}\sum_{i=1}^n\log \mathcal{F}_{2}  \left(\frac{\mu^2}{s_{ii+1}},g_\varepsilon,\varepsilon\right) 
+
r_{n}\big(\{p_i\},g_\varepsilon\big) 
+
O(\varepsilon)
,
\end{align}
where the on-shell Sudakov form factor $\mathcal{F}_{2}$ has the following all-order expansion 
\cite{Mueller:1979ih,Magnea:1990zb,Sterman:2002qn} 
\begin{align}
\label{FF2}
\log \mathcal{F}_{2} \left(\frac{\mu^2}{s},g_\varepsilon,\varepsilon\right)
=
-
\frac{1}{2}\sum_{L=1}^{\infty} g_\varepsilon^{2 L}
\left[ \frac{\Gamma^{(L)}_{\rm cusp}}{(L\varepsilon)^2}
+
\frac{G^{(L)}}{(L\varepsilon)}
+
c^{(L)}\right]
\left(- \frac{\mu^2}{s}\right)^{L\varepsilon}
+
O(\varepsilon)
,
\end{align}
in terms of the $D$-dimensional 't Hooft coupling $g^2_\varepsilon \equiv g^2 (4 \pi {\rm e}^{- \gamma_{\rm E}})^{\varepsilon}$. Here, 
the double and single poles are accompanied by two infinite sets of coefficients, defining the cusp $\Gamma_{\rm cusp} (g)$ and 
collinear $G (g)$ anomalous dimensions,
\begin{align}
\label{GCuspAndGPT}
\Gamma_{\rm cusp}(g)
&=
\sum_{L =1}^{\infty}\Gamma_{\rm cusp}^{(L)} g^{2 L}
=
4 g^2 - 8\zeta_2 ^4 g^4 + 88 \zeta_4 g^6 + \ldots
\, , \nonumber\\
G(g)
&=
\sum_{L=1}^{\infty}G^{(L)} g^{2 \ell}
=
-
4 \zeta_3 g^4+ \left( 32 \zeta_5+\ft{80}{3}\zeta_2\zeta_3 \right) g^6 + \ldots
\, .
\end{align}
The representation \re{FF2} displays a very profound interactive structure of perturbation theory, which was highly predictive within 
the context of the $\mathcal{N}=4$ sYM. The entire kinematical dependence of the Sudakov form factor 
\begin{align}
\label{BDSExpDimReg}
\log \mathcal{F}_{2}^{\rm BDS}
\left(\frac{\mu^2}{s},g_\varepsilon,\varepsilon\right)
=
\sum_{L=1}^{\infty}g_\varepsilon^{2L}
[\Gamma_{\rm cusp}^{(L)} + L \varepsilon G^{(L)} + (L \varepsilon)^2 c^{(L)} + O (\varepsilon^3)] 
\mathcal{F}_{2}^{(1)} \left(\frac{\mu^2}{s}, L\varepsilon\right)
\, ,
\end{align}
is incorporated through its one-loop expression
\begin{align}
\mathcal{F}_{2}^{(1)} \left(\frac{\mu^2}{s},\varepsilon\right)
= 
- 2 \frac{{\rm e}^{\varepsilon \gamma_{\rm E}} \Gamma^2 (-\varepsilon) \Gamma (1+\varepsilon)}{ \Gamma (1-2\varepsilon)} \left(- \frac{\mu^2}{s} \right)^{\varepsilon} 
\end{align}
with the loop order $L$ merely rescaling the regularization parameter $\ep$, $\ep \to L \ep$. This is known as the BDS ansatz \cite{BernBDS}. 
The non-trivial behavior of $\mathcal{F}_{n}$ as adjacent particle legs become collinear, i.e., $p_i \to z (p_i + p_{i+1})$ and $p_{i+1} \to (1-z) 
(p_i + p_{i+1})$, is entirely encoded in $\mathcal{F}_{2} ^{\rm BDS}$. The finite remainder function $r_{n}\big(\{p_i\},g_\varepsilon \big)$ includes 
deviations from the BDS functional form for form factors with $n>2$ legs. It is customary to cancel out the BDS ansatz from the full form
factor and define a finite remainder function $\mathcal{R}_n$ as
\begin{align}
\label{Remainder}
\mathcal{R}_n = \log \left( \mathcal{F}_{n} /\mathcal{F}_{n}^{\rm BDS} \right)
\, , 
\end{align}
where $\mathcal{F}_{n}^{\rm BDS}$ is a generalization of Eq.\ \re{BDSExpDimReg} where subscript 2 gets replaced by $n$. For instance,
at the two-loop order, it reads
\begin{align}
\label{IterativeR2}
\mathcal{R}_n^{(2)} (\varepsilon)
=
\mathcal{F}_{n}^{(2)} (\varepsilon) 
- 
\ft12 \left( \mathcal{F}_{n}^{(1)} (\varepsilon) \right)^2
-
[\Gamma_{\rm cusp}^{(2)} + 2 \varepsilon G^{(2)} + (2 \varepsilon)^2 c^{(2)} ] \mathcal{F}_{n}^{(1)} (2 \varepsilon) + O (\varepsilon)
\, ,
\end{align}
and we dropped kinematical arguments in all functions to accentuate the iteration. The so-defined remainder is endowed with trivial 
collinear limits, i.e., it vanishes or, at most, becomes a (transcendental) constant.

The near-mass-shell behavior for the Coulomb branch form factors for $n=2$ at three loops and $n=3$ (but with the non-minimal operator
$\tr\phi_{12}^2$) at two loops was studied by us in Refs.\ \cite{Belitsky:2022itf,Belitsky:2023ssv} and \cite{Belitsky:2024agy}, respectively. 
These considerations can be recast as the following conjectural all-order expression
\begin{align}
\label{AmplFactorCB}
\log \mathcal{F}_{n} = \frac{1}{2} \sum_{i=1}^n \log \mathcal{F}_{2} \left(\frac{m}{s_{ii+1}},g\right) 
+ 
r_{n}\big(\{p_i\},g\big) 
+
\ft{n}2 D (g) + O(m)
\, .
\end{align}
This is to be contrasted with the one for the dimensionally regularized conformal theory since the two-leg Sudakov form factor
is given now by an exact formula \cite{Belitsky:2022itf}
\begin{align}
\label{LogM2octagon}
\mathcal{F}_{2} \left(\frac{m}{s},g\right) = -\frac{\Gamma_{\rm oct} (g)}{2} \log^2 \left(- \frac{m}{s}\right) - D(g)+\mathcal{O}(m),
\end{align}
in terms of the octagon anomalous dimension $\Gamma_{\rm oct} (g)$ and a finite part $D (g)$
\begin{align}
\label{ExactGammaOct}
\Gamma_{\rm oct}(g)
&
= 
\frac{2}{\pi^2}\log  \cosh \left(2 \pi g \right) = - 4 g^2 + 16 \zeta_2 g^4
+ \ldots
\, , \\
D(g)
&
=
\frac{1}{4}\log \frac{\sinh( 4 \pi g )}{4\pi g} = 4 \zeta_2 g^2 - 32 \zeta_4 g^4
+ \ldots
\, .
\end{align}
These emerged in the context of correlation functions of infinitely heavy half-BPS operators $\mathcal{O} |_{n \to \infty}$ 
\cite{Coronado:2018cxj} and were calculated exactly in Refs.\ \cite{Belitsky:2019fan,Belitsky:2020qzm}. Comparing the 
one-loop results for the on- and off-shell form factors, one uncovers the well-known factor-of-two-difference for the coefficient
of the logarithmic behavior, known for almost half a century \cite{Sudakov:1954sw,Jackiw:1968zz}. It was attributed to 
the existence of an extra regime of loop momenta in quantum corrections  \cite{Fishbane:1971jz,Mueller:1981sg,Korchemsky:1988hd} 
dubbed ultrasoft in modern literature. 
In Abelian theories, this was the end of it since the infrared logarithms exponentiate in terms of the one-loop exact bremsstrahlung 
function as a reflection of the Poissonian nature of photon emissions. In non-Abelian theories, the above result demonstrates that 
infrared behavior for the on- and off-shell regimes is governed by different functions of the coupling, $\Gamma_{\rm cusp}$ and
$\Gamma_{\rm oct}$, respectively. Another distinguished feature of the off-shell result is the absence of an analogue for the
collinear anomalous dimension $G (g)$. It was demonstrated quite recently \cite{Belitsky:2024yag} that this is a consequence of 
subtle cancellations between the ultrasoft and collinear regimes of loop momentum integrals in the full form factor. Last but
not least, the finite parts in Eqs.\ \re{AmplFactor} and \re{AmplFactorCB} are different in the two cases, as was confirmed in 
Ref.\ \cite{Belitsky:2024agy}, and the iterative structure of higher-order correction is lost, in particular, there are no immediate 
generalizations of (\ref{Remainder}) and (\ref{IterativeR2}).

After providing these general comments, we start with an explicit two-loop analysis of the minimal $n=3$ form factor first.
This will be followed by results for any $n>3$.

\section{Uplift}
\label{UpliftSection}

As we discussed at length in the introduction, the $\mathcal{N} = 4$ sYM on the Comlomb branch can be obtained by a generalized 
dimensional reduction, which keeps out-of-four-dimensional momenta nonvanishing. A particularly beneficial starting point is 
the maximally supersymmetric $\mathcal{N} = (1,1)$ sYM in six dimensions \cite{Bern:2010qa}. Its on-shell physics can be recast 
in terms of unconstrained spinor-helicity formalism \cite{Cheung:2009dc} which receives a superspace generalization as well 
\cite{Dennen:2009vk}. This approach was used to reconstruct $D$-dimensional integrands for four-dimensional external states 
\cite{Bern:2010qa}. However, one can equally well use it to find four-dimensional integrands of $D$-dimensional external states. 
For a low number of legs, this approach demonstrates that a naive uplift of four-dimensional integrands to $D$-dimensions is 
legitimate since integrands do not depend explicitly on the number of space dimensions $D$. The latter enters implicitly through Lorentz 
products only. So, one can interpret these in any number of dimensions: the integrands of the $\mathcal{N} = 4$ sYM look the same. 
This naive uplift is expected to start failing from a certain critical number of external legs. The presence of the so-called $\mu$-terms 
is a sign of this, as is the case for hexagon amplitudes \cite{Bern:2008ap}.

\begin{figure}[t]
\begin{center}
\mbox{
\begin{picture}(0,280)(150,0)
\put(0,0){\insertfig{10}{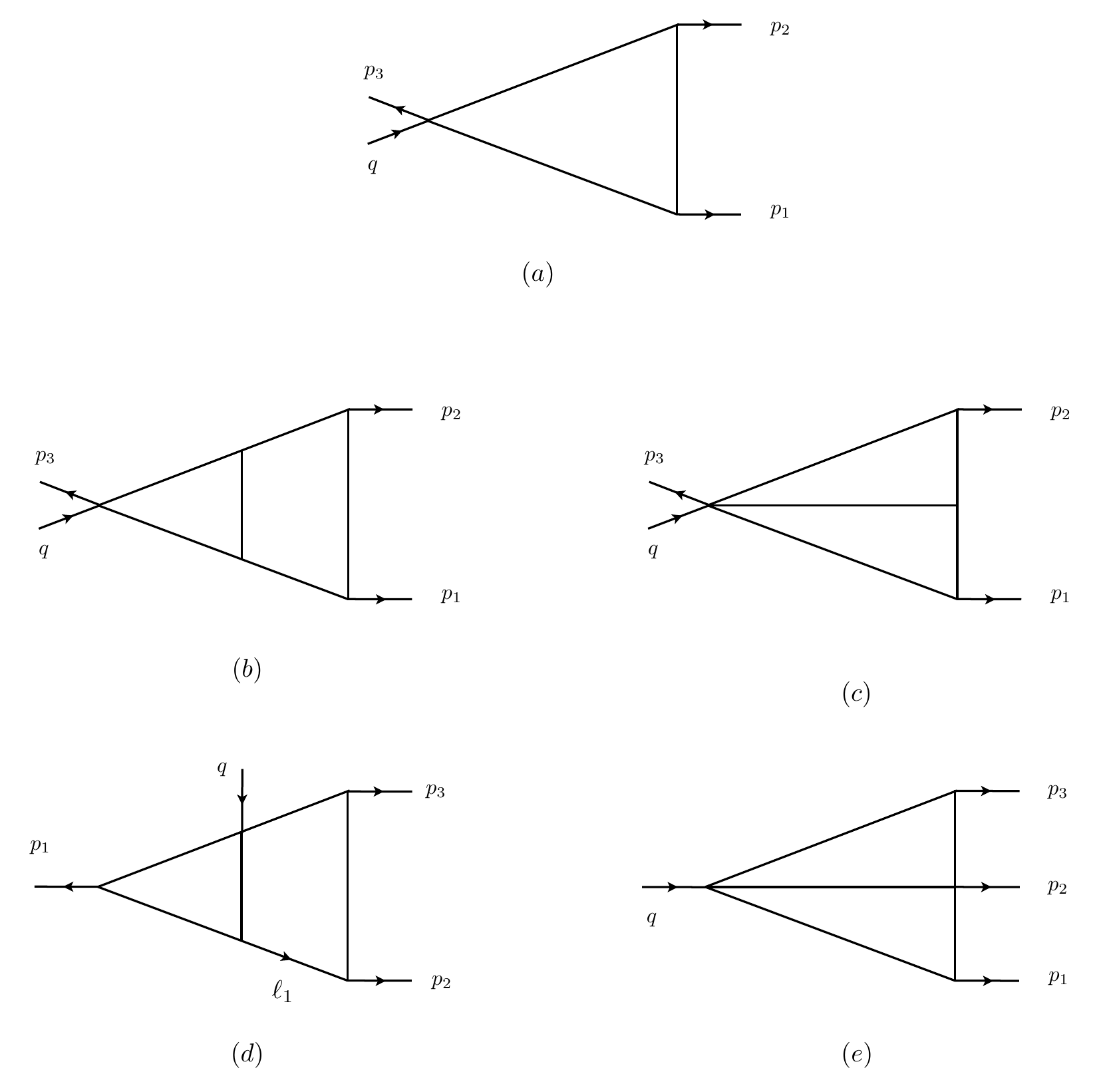}}
\end{picture}
}
\end{center}
\caption{\label{fig2loop} One and two-loop graphs contributing to three particle form factor: a) is ${\rm Tri} (p_1, p_2)$, b) is ${\rm DTri}_1 
(p_1, p_2)$, c) is ${\rm DTri}_2 (p_1, p_2)$, d) is ${\rm TriBox}  (p_2, p_3,p_1)$, and e) is ${\rm TriBox}_{\rm red} (p_1, p_2, p_3)$.}
\end{figure}
 
For the $n=3$ form factor, the $D$-dimensional integrand at two-loop order\footnote{Let us point out that in the conformal case,
this form factor is currently known up to the sixth order of perturbative series from the functional bootstrap that relies on the use of 
analyticity and integrability \cite{Basso:2024hlx}. Its three-loop integrands were constructed earlier in Refs.\ \cite{Lin:2021kht,Lin:2021qol} 
and recently integrated explicitly in Ref.\ \cite{Henn:2024pki}.} was constructed in Ref.\ \cite{Brandhuber:2014ica} making 
use of the four-dimensional unitarity cut technique in terms of a set of scalar integrals. In fact, all of them are formally dual-conformally 
covariant. This is our starting point in this section. We promote all Lorentz products of the loop $\ell$ and external $p$ momenta, 
cumulatively called $k=(\ell, p)$, in their integrands upwards to six dimensions
\begin{align}
k_i \cdot k_j \to K_i \cdot K_j
\, ,
\end{align}
and then pass to the region's momenta 
\begin{align}
P_i = X_i - X_{i+1}
\, ,
\end{align}
to {\sl enforce the six-dimensional momentum conservation}. The required kinematical conditions are then implemented by considering
the out-of-four-dimensional components $y_i$ of $X_i = (x_i, y_i)$ as null, i.e., $y_i^2 = 0$, so that all internal lines remain massless 
from the four-dimensional point of view. The six-dimensional external null momenta $P_i^2 = 0$ on the other hand generate nonvanishing 
four-dimensional virtualities provided $y_i \cdot y_j \neq 0$ ($i \neq j$) such that $p_i^2 = y_{ii+1}^2$. The latter becomes the infrared 
regulator of loop momenta so that the measure can safely be reduced down to four dimensions.

\begin{figure}[t]
\begin{center}
\mbox{
\begin{picture}(0,85)(245,0)
\put(0,0){\insertfig{17}{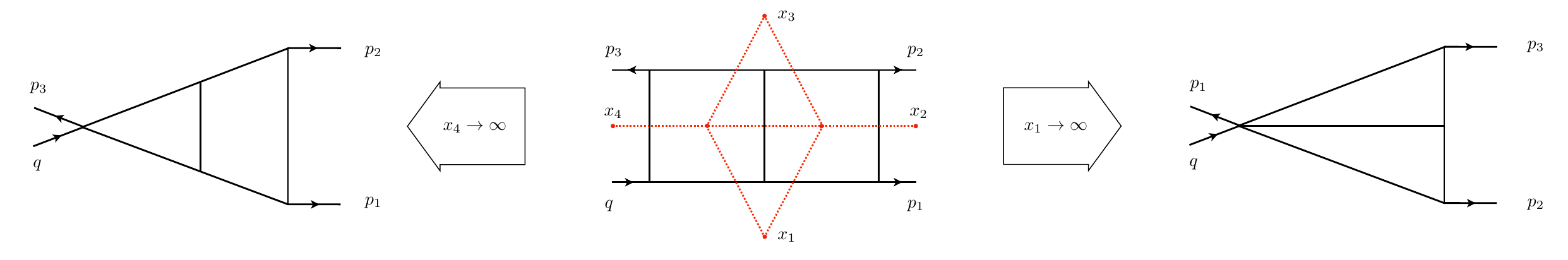}}
\end{picture}
}
\end{center}
\caption{\label{DCrelFig} Double triangle integrals DTri as limits of the off-shell double box.}
\end{figure}

The kinematics for the $n=3$ form factor is formally equivalent to the massive four-leg amplitude with the momentum conservation
condition $p_1 + p_2 + p_3 - q = 0$ such that the Madelstam invariants $s_{ii+1}$ obey the familiar relation
\begin{align}
s_{12} + s_{23} + s_{13} = q^2 - 3 m
\, ,
\end{align}
making use of the conventions introduced earlier in Eq.\ \re{KinInvariants}. We will be interested in this work only in the near mass-shell
limit when $m \to 0$. In these circumstances, it is convenient to adopt the same variables as were used in the conformal case 
\cite{Brandhuber:2014ica} for easier comparison. Namely
\begin{align}
\label{mandelstam2}
u = \frac{s_{12}}{q^2}
\, ,\qquad 
v =\frac{s_{23}}{q^2}
\, ,\qquad
w=\frac{s_{13}}{q^2}
\, ,\qquad
u + v + w \simeq 1
\, ,
\end{align}
where we ignored power-suppressed corrections on the right-hand side of the linear dependence equation. With these conventions,
the one- and two-loop contributions to the form factor read \cite{Brandhuber:2014ica}
\begin{align}
\label{1loopFF}
\mathcal{F}_3^{(1)}
= 
\sum_{k = 0}^2 \mathbb{P}^k \Big[s_{12} {\rm Tri} (p_1, p_2) \Big]
\, ,
\end{align}
and
\begin{align}
\label{2loopFF}
\mathcal{F}_3^{(2)}
=
\sum_{k = 0}^2 \mathbb{P}^k
\Big[
&
s_{12}^2 {\rm DTri}_1 (p_1, p_2) + s_{12}^2 {\rm DTri}_2 (p_1, p_2) 
\\
&
+
s_{23} {\rm TriBox}  (p_1, p_2, p_3)
+ 
s_{12} {\rm TriBox}  (p_3, p_2,p_1) 
- 
q^2 {\rm TriBox}_{\rm red} (p_1, p_2, p_3)
\Big]
\, , \nonumber
\end{align}
respectively. Here $\mathbb{P}$ is the shift operator that changes the particle momentum $p_i$ label $i$ by $+1$, $\mbox{mod}(3)$. 
For instance, $\mathbb{P}^2\log (s_{12}/s_{23})=\log (s_{31}/s_{12})$. All Feynman integrals admit the same form
\begin{align}
{\rm Int}_i = \int \prod_j \frac{d^2 \ell_j}{i \pi^2} \frac{{\rm num_i}}{{\rm denom_i}}
\, ,
\end{align}
where the denominators can be read off from the graphs in Fig.\ \ref{fig2loop}. All numerators are one except for the graph in 
Fig.\ \ref{fig2loop} (d), which is ${\rm num_{(d)}} = (\ell_1 + p_1)^2$. The one-loop triangle graph is given by the standard 
$\Phi_1$ function 
\begin{align}
s_{12} {\rm Tri} (p_1, p_2) = \Phi_1 \left( - \frac{m}{s_{12}}, - \frac{m}{s_{12}} \right)
\, ,
\end{align}
of Davydychev and Ussyukina (DU) \cite{Usyukina:1992jd,Usyukina:1993ch},
\begin{align}
\label{BoxFunct}
\Phi_\ell(x,y)=-\sum_{j=\ell}^{2\ell}\frac{j!(-1)^{j}\log^{2 \ell - j}\left( \frac{y}{x} \right)}{\ell! (j - \ell)! (2 \ell - j)!}
\frac{1}{\lambda}\left[ \mbox{Li}_{j} \big(-(\rho x)^{-1}\big) - \mbox{Li}_{j}\big( -(\rho y)^{+1}\big) \right]
\, ,
\end{align}
where $\rho$ and $\lambda$ are functions of $x$ and $y$,
\begin{align}
\lambda(x,y)=[(1-x-y)^2-4xy]^{1/2} \, , \qquad \rho(x,y)=2 [1-x-y-\lambda(x,y)]^{-1} \, ,
\end{align}
The variables $x$ and $y$ are the conformal cross ratios of external kinematical variables. At two loops, despite apparent differences 
between topologies of the graphs (b) and (c) in Fig.~\ref{fig2loop}, they are expressed by the same double box DU function $\Phi_2$ 
as can be seen from their limiting representation in dual variable space as was observed in Ref.\ \cite{Bork:2010wf} and demonstrated 
in Fig.~\ref{DCrelFig},
\begin{align}
s_{12}^2 {\rm DTri}_1 (p_1, p_2)
=
s_{12}^2 {\rm DTri}_2 (p_1, p_2)
=
\Phi_2 \left( - \frac{m}{s_{12}}, - \frac{m}{s_{12}} \right)
\, .
\end{align}
Therefore, the only integrals that need separate attention are the ones in Fig.\ \ref{fig2loop} (d) and (e). 

\section{Generalization to any $n$}

The uplift at the two-loop level can be executed for $n>3$ form factor integrands \cite{Brandhuber:2014ica} as well. One loop
does not contain functionally new contributions except for adding additional permutations
\begin{align}
\label{1loopFFNp}
F_n^{(1)}= \sum_{k = 0}^{n-1} \mathbb{P}^k \Big[ s_{12} {\rm Tri} (p_1, p_2) \Big]
\, ,
\end{align}
where now $\mathbb{P}$ is the shift operator that shifts all labels of particle momenta $p_i$ by $+1$, $\mbox{mod}(n)$. While at two 
loops, 
\begin{align}
\label{2loopFFNp}
F_n^{(2)}
=
&
\sum_{k = 0}^{n-1} \mathbb{P}^k
\Big[
s_{12} {\rm DTri}_1 (p_1, p_2) + s_{12} {\rm DTri}_2 (p_1, p_2)
\\
& +  s_{23} {\rm TriBox}  (p_1, p_2, p_3) + s_{12} {\rm TriBox}  (p_3, p_2, p_1) +
\nonumber\\
&
- q^2 {\rm TriBox}_{\rm red} (p_1, p_2, p_3)
+\frac{1}{2}\sum_{j=3}^{n}s_{12}s_{jj+1}~{\rm Tri} (p_1, p_2){\rm Tri} (p_j, p_{j+1})  \Big]
\, , \nonumber
\end{align}
there is a new term stemming from the interaction of fields in two independent pairs of lines (the last term above).
Before we turn to the calculation of contributing integrals, let us reproduce the above expressions using a complementary 
technique.

\section{$\mathcal{N} = 1$ superspace calculation}
\label{sec:part4}

We are going to perform our analysis making use of the $\mathcal{N} = 1$ superspace formulation of the $\mathcal{N} = 4$ sYM. We 
revisit the consideration in \cite{Bork:2010wf}, where the two-loop form factor $\mathcal{F}_n$ was addressed for the first time
in the conformal phase, and generalize it to the Coulomb branch. Along the way, we will correct some inaccuracies percolating 
that work, making them agree with Ref.\  \cite{Brandhuber:2014ica}. We will discuss only the near-mass-shell limit within this 
formalism rather than full-fledged off-shell kinematics, which requires dedicated studies to properly treat effects due to gauge 
symmetry breaking.

The $\mathcal{N} = 1$ formulation of the maximal supersymmetry in four dimensions lacks SU$_{\rm R}$(4) covariance since it splits 
fermions and scalars between three Wess-Zumino multiplets, their conjugates, and a vector multiplet. They are encoded, respectively, 
in three chiral $\Phi_i$ and anti-chiral $\bar\Phi_i$ superfields and a real vector superfield $V$. The lowest components of the chiral fields 
in the Grassmann expansion are certain components of the scalar sextet, namely,
\begin{align}
\Phi^i |_{\theta = 0} = \bar\phi^{i4}
\, , \qquad
\bar\Phi_i |_{\bar\theta = 0} = \phi_{i4}
\, .
\end{align}
These obey the reality condition $\bar\phi^{i4} = \ft12 \varepsilon^{ijk}\phi_{jk}$. Only the SU$_{\rm R}$(3) subgroup of the R-symmetry group
is left manifest. However, this will not bear any significance on our calculation. As always, the advantage of the superspace formulation is
an automatic sum over the plethora of states propagating in quantum loops.

\begin{figure}[t]
\begin{center}
\mbox{
\begin{picture}(0,70)(90,0)
\put(0,0){\insertfig{6}{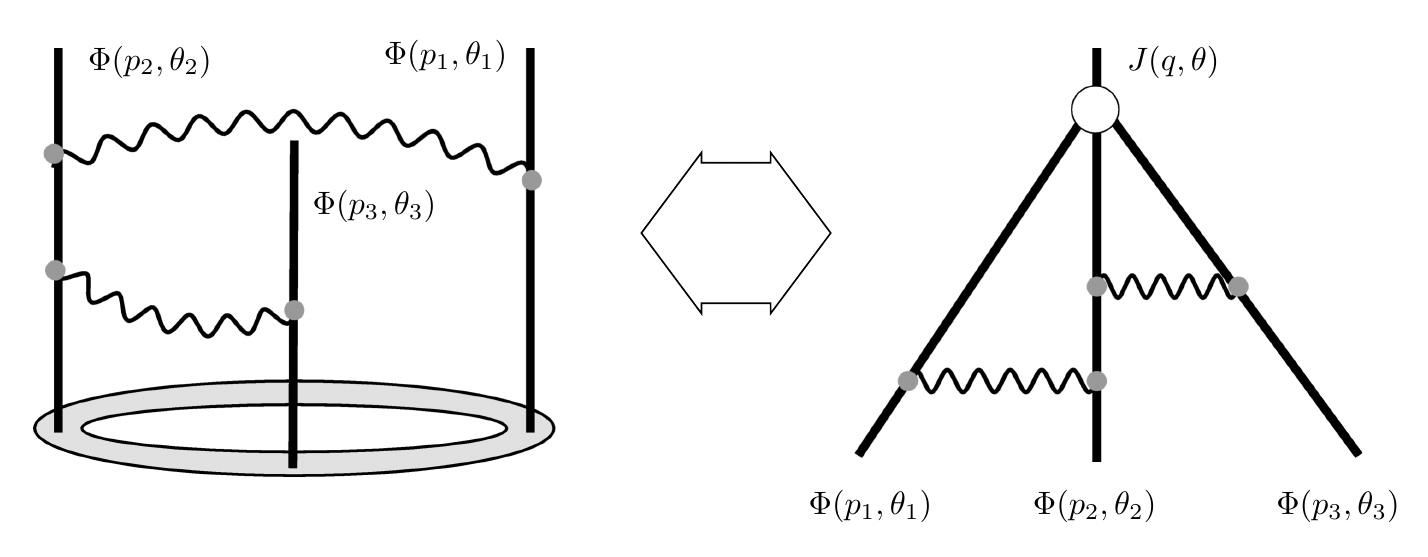}}
\end{picture}
}
\end{center}
\caption{\label{figSuer0} The equivalence of single-trace superfield Feynman diagrams (left) to traditional ones (right). The white blob in the
right panel stands for the operator vertex ${\rm tr}\Phi^n$ (with n=3) and is shown by the annulus on the left-hand side. The solid/wavy 
lines represent chiral/vector propagators.}
\end{figure}

To start with, we add the super-space generalization of the operator \re{phi12n},
\begin{align}
\tr \phi_{12}^n
\to 
\tr (\Phi^3)^n
\,
\end{align}
to the sYM action with a supersource $J$. In fact, we will use a shorthand notation $\Phi$ for $\Phi^3$ to avoid cluttering formulas with
unnecessary indices. Thus,
\begin{align}
S_{\mathcal{N} = 4} \mapsto S_{\mathcal{N} = 4} + \int d^6 z J(z) \tr \Phi^n
\, .
\end{align}
where the $\mathcal{N}$=4 sYM action in terms of the $\mathcal{N}$=1 superfields is given by \cite{Grisaru:1980jc,Gates:1983nr}
\begin{align}
\label{Sn4n1}
S_{N=4} 
&
= 
\frac{1}{8 g^2_{\scriptscriptstyle\rm YM}}\int d^6z \tr (W^\alpha W_\alpha) 
\\
&+
2 \int d^8z \tr \left(e^{-g_{\scriptscriptstyle\rm YM}V} \bar\Phi_i e^{g_{\scriptscriptstyle\rm YM}V} \Phi^i \right) 
+
\ft{i}{\sqrt{2}}g_{\scriptscriptstyle\rm YM}\int 
\tr \left(d^6z \Phi^1[\Phi^2,\Phi^3] - d^6\bar{z} \bar\Phi_1[\bar\Phi_2,\bar\Phi_3]\right)
\, . \nonumber
\end{align}
The kinetic term for the vector field is expressed here via the superfield strength $W_\alpha= \ft14 \bar{D}^2
(e^{-2 g_{\scriptscriptstyle\rm YM}V}D_\alpha e^{2 g_{\scriptscriptstyle\rm YM}V})$ and the integration measures are
$d^6z=d^2\theta d^4x$, $d^6\bar{z}=d^2\bar\theta d^4x$, and  $d^8z=d^2\theta d^2\bar{\theta} d^4x$. Passing from the 
generating functions of connected Green functions $i W = \log Z$ via the Legendre transform with respect to a 
supersource for the chiral superfield $\Phi$, we get the textbook effective action for their expectation values 
$\Gamma [\Phi_{\rm cl}, J]$. We need to keep track of only one term in its expansion with the linear dependence on the 
source $J$ and $n$ powers of $\Phi_{\rm cl}$,
\begin{align}
\label{GammaEffectiveLocal}
\Gamma [\Phi_{\rm cl}, J]
= \int d^2 \theta d^4 p_1 \ldots d^4p_n \mathcal{M}_n (p_1,{\dots},p_n)
J(q,\theta) \tr (\Phi_{\rm cl}(p_1,\theta){\dots}\Phi_{\rm cl}(p_n,\theta)) 
+ \dots
\, .
\end{align}
The perturbative calculation of $\mathcal{M}_n$ is equivalent to the loop expansion of the minimal form factor $\mathcal{F}_n.$

For the lowest two orders of the 't Hooft expansion,
\begin{align}
\mathcal{M}_n = 1 + g^2 \mathcal{M}_n^{(1)} + g^4 \mathcal{M}_n^{(2)} + O (g^6)
\, ,
\end{align} 
the number of contributing supergraphs is relatively small and the diagrammatic calculation is manageable \cite{Bork:2010wf}. 
This setup also allows us to investigate both the on- and near-mass-shell behavior of the amplitude $\mathcal{M}_n$. We will 
use a slightly unorthodox presentation for them, designating the single-trace operator by an annulus with fields emanating
from it, reflecting its cyclicity. The correspondence with more traditional rules is exemplified in Fig.\ \ref{figSuer0}. 

\begin{figure}[t]
\begin{center}
\mbox{
\begin{picture}(0,145)(130,0)
\put(0,0){\insertfig{9}{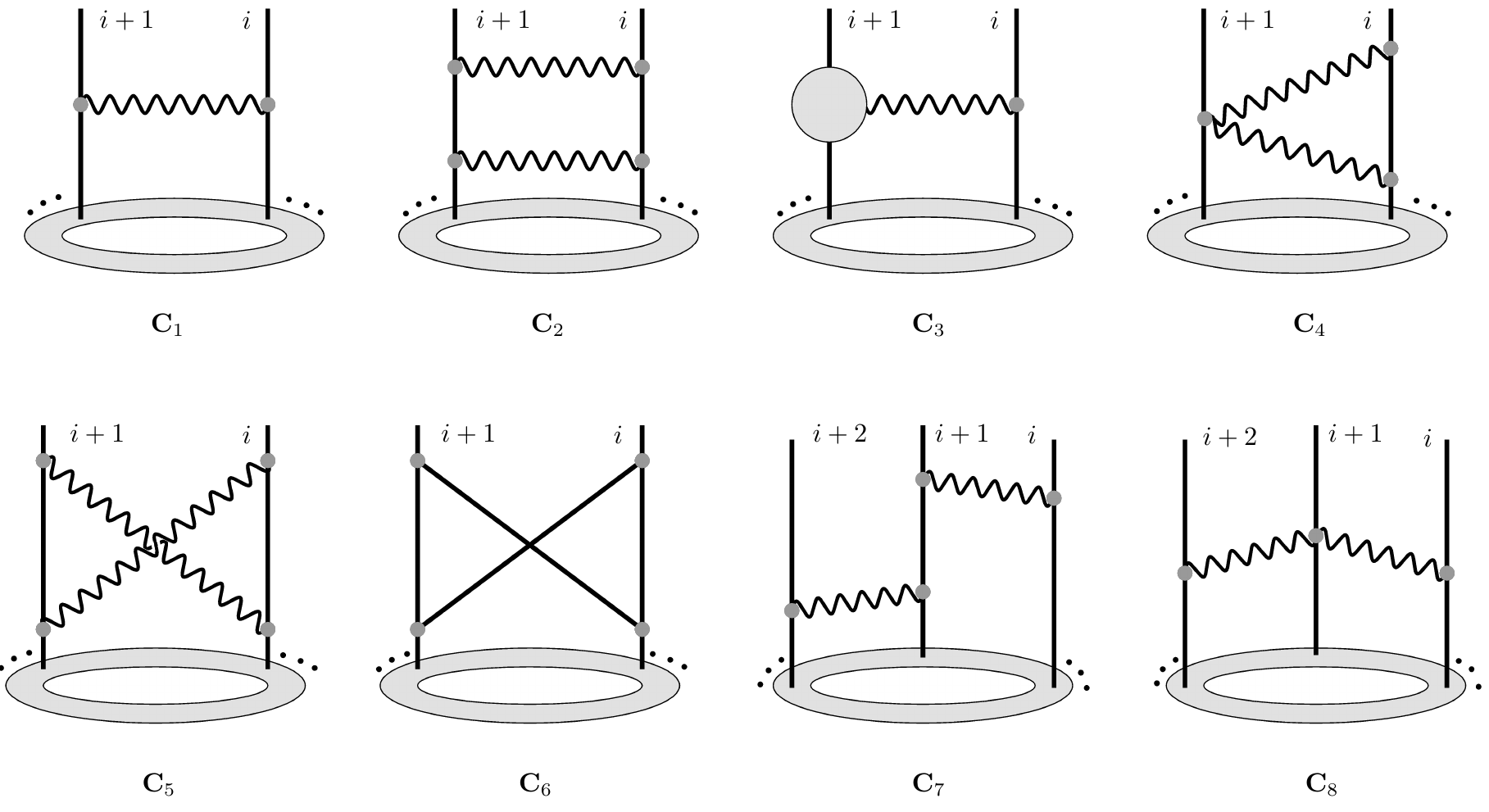}}
\end{picture}
}
\end{center}
\caption{\label{figSuer1}  Supergraphs contributing to $\tr \Phi^n$ form factor. For $n \geq 3$, the supergraphs $\textbf{C}_{5,6}$ are 
absent in the planar limit. For $n=2$, the supergraphs $\textbf{C}_{7,8}$ do not exist. The grey blob in ${\bf C}_{3}$ corresponds to 
the one-loop effective vertex, see, e.g., \cite{Grisaru:1980jc}. The explicit position of $D$-derivatives is not displayed in any of the graphs.}
\end{figure}
 
At one and two loops, contributing families of supergraphs are shown in Fig.\ \ref{figSuer1},
\begin{align}
\label{M1and2}
{\bf C}_1 + \mbox{perm.} \to \mathcal{M}_n^{(1)}
\, , \qquad
\sum_{i = 2}^8 {\bf C}_i + \mbox{perm.} \to \mathcal{M}_n^{(2)}
\, ,
\end{align}
where $\mbox{perm.}$ stands for cyclic permutations.
Up to now, we have not specified whether we are discussing an on- or off-shell regime. This choice has to be made, however, at the 
time of implementing the algebra of super-covariant derivatives and corresponding reductions. If the fields are treated as on-shell, we 
impose the equation of motion $\bar D^2 D^2 \Phi_{\rm cl}=0$ on the external `classical' field as well as the on-shell condition 
$p_i^2=0$ on corresponding external momenta. For the off-shell situation, the superfields $\Phi_{\rm cl}$ are left unconstrained and we 
keep $p_i^2\neq 0$ for external momenta. However, the latter will be chosen near their mass shell. This is done to avoid dealing
with the off-shellness in the numerators of Feynman supergraphs. Of course, there is a possibility that $p_i^2$ will get accompanied by
a singular momentum integral $I$ that yields a power-enhanced effect after momentum integrations, i.e., $I \sim 1/p_i^2$. However,
we observed no such contributions in the two-loop order. This is in accord with a similar observation at one-loop order for off-shell
superspace amplitudes \cite{Santini:2012uda}.

\begin{figure}[t]
\begin{center}
\mbox{
\begin{picture}(0,283)(230,0)
\put(0,0){\insertfig{16}{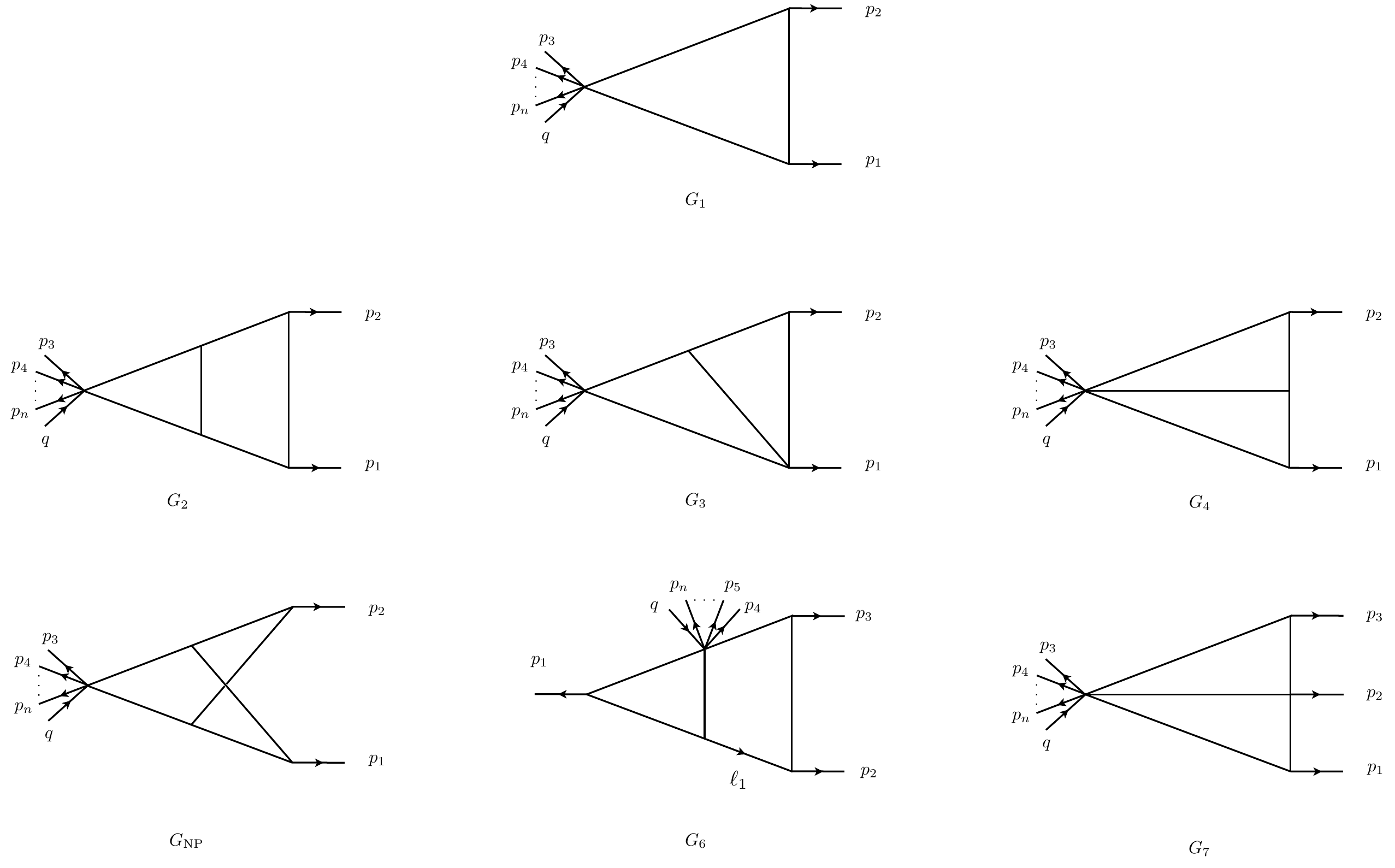}}
\end{picture}
}
\end{center}
\caption{\label{figSuer2}  Scalar integrals originating from supergraphs for the general $n$ case. As before arrows on the lines with momenta $l$ and $p$ corresponds to the presence of numerator $(p_1+\ell_1)^2$.}
\end{figure}

Now, we are in a position to perform the algebra of super-derivatives and integral reduction. We will focus on the $n=2$ and $n=3$ cases
and discuss them in parallel since we will draw important conclusions from the former case for the latter. For these two cases, some of the
graphs in Fig.\ \ref{figSuer1} are missing due to the color flow and/or operator field multiplicity. At one loop, they receive contributions from 
the same graph topology ${\bf C}_1$ as was already shown in Eq.\ \re{M1and2}. At the two-loop level, we have
\begin{align}\label{phi2super}
\sum_{i=2}^4 {\bf C}_i +{\bf C}_5 + {\bf C}_6 + \mbox{perm.} \to \mathcal{M}^{(2)}_2
\, ,
\end{align}
for the $\tr \Phi^2$ operator, and
\begin{align}\label{phi2super1}
\sum_{i=2}^4 {\bf C}_i
+ \left. {\bf C}_7 + {\bf C}_7 \right|_{p_i \leftrightarrow p_{i+2}} + {\bf C}_8 + \mbox{perm.}
\to \mathcal{M}^{(2)}_3 
\, ,
\end{align}
for $\tr \Phi^3$. The combinatorics and algebra of super-derivatives are identical among all two-particle ${\bf C}_1$ and ${\bf C}_{2,3,4}$ 
supergraphs. The three-particle supergraphs ${\bf C}_{7,8}$ that involve simultaneous interaction of three legs are impossible in the $n=2$ 
case. On the other hand, ${\bf C}_{5,6}$ are absent for $n=3$ in the planar limit. This becomes obvious once the color structure of
the operator vertex is accounted for. For the $n=2$ operator, the latter is proportional to $\delta^{ab}$ and allows for non-planar 
topologies even in the leading order in color. Thus, cross-propagator scalar integrals contribute even in the planar limit. For the $n=3$ 
operator, its vertex is proportional to the $f^{abc}$-structure constants, and nonplanar graphs are power-suppressed. The same occurs 
for $n>3$.

The algebra of super-covariant derivatives is identical in the on-shell and near-mass-shell, i.e., $p_i^2= - m \to 0$, situations. The former 
is rather elementary for all supergraphs except ${\bf C}_{5,6}$.  All distinct topologies that can emerge from algebraic manipulations are 
shown in Fig.\ \ref{figSuer2}. If not for $G_3$ they coincide with the integrals in Fig.\ \ref{fig2loop}, namely $G_1 = {\rm Tri}$, 
$G_2 = {\rm DTri}_1$,  $G_4 = {\rm DTri}_2$, $G_6 = {\rm TriBox}_{\rm red}$, and $G_7 = {\rm TriBox}$. $G_{\rm NP}$ is intrinsic to the 
two-leg form factor only, while $G_3$ does not appear from the unitarity-cut construction since the latter has the {\sl no-triangle rule} in 
dressing on-shell external legs built-in automatically in the formalism, as pointed out in \cite{Brandhuber:2014ica}. Notice that $G_4$ 
has the very same topology as $G_3$ but is allowed since triangles include the off-shell vertex with the $q$-momentum. 

\begin{table}[t]
\begin{center}
\begin{tabular}{|c|c|c|}   
\hline
$\mbox{supergraphs}$ & $\tr \Phi^2$ & $\tr\Phi^3$\\
\hline
$\textbf{C}_1$ & $q^2 G_1$ & $s_{12} G_1$\\
\hline
$\textbf{C}_2$ & $q^4 G_2$ & $s_{12}^2 G_2$\\
\hline
$\textbf{C}_3$ & $q^2 (G_3+G_4)$ & $s_{12} (G_3+G_4)$\\
\hline
$\textbf{C}_4$ & $-q^2G_3$ & $-s_{12} G_3$\\
\hline
$\textbf{C}_5+\textbf{C}_6$ & $x_1 q^2 G_3+x_2 q^2 G_4 + x_3G_{\rm NP}$ &$0$\\
\hline
$\textbf{C}_7$ &$0$ &$s_{23} G_{6}$\\
\hline
$\textbf{C}_8$ &$0$ &$q^2 G_7$\\   \hline
\end{tabular}
\caption{Contributions to the form factors of the $\tr\Phi^2$ and $\tr\Phi^3$ operators from individual supergraphs.}
\label{chir}
\end{center}
\end{table}

The results for the reduction of the supergraphs to the Feynman integrals $G_i$ are summarized in Table \ref{chir}. For 
${\bf C}_{5}+{\bf C}_{6}$, we devised a shortcut though. Namely, we wrote the most general linear combination of scalar integrals 
accompanied by unknown coefficients $x_i$. Then, we fixed the aforementioned coefficients by matching the total sum 
graphs 
\begin{align}
\label{xCond1}
\mathcal{M}_{2}^{(2)}
\equiv q^4 G_2+q^2 (G_3+G_4) - q^2G_3+x_1 q^2~G_3+x_2 q^2 G_4+x_3 G_{\rm NP}
=
q^4 G_2 + \ft{1}{4} G_{\rm NP}
\, ,
\end{align}
to the available results in the literature \cite{vanNeerven:1985ja,Gehrmann:2011xn}, shown after the second equality sign.
This provides us with an unambiguous solution for the coefficients, namely, $x_1=0$, $x_2=-1$, and $x_3=\ft14$. 

Let us provide a few comments on our findings. The relative coefficients for integrals as they emerge from supergraphs ${\bf C}_1$, 
${\bf C}_2$, ${\bf C}_3$, and ${\bf C}_4$ are identical among all the $\tr \Phi^n$ form factors, i.e., they are universal. The relative 
sign and coefficient between $G_3$ and $G_4$ in ${\bf C}_3$ supergraphs are completely fixed by the algebra of super-derivatives. 
With these results at hand, we observe the much-needed cancellation of the $G_3$ integral in the $\tr \Phi^n$ for all $n>3$ in 
agreement with the unitarity technique. This corrects the analysis of Ref.\ \cite{Bork:2010wf}. We obtain, for $n=3$ at two loops
\begin{align}
\label{xCond2}
\mathcal{M}_{3}^{(2)}
\equiv s_{12}^2 G_2 + s_{12} G_4 + s_{23} G_6 + \left. s_{23} G_6 \right|_{p_1 \leftrightarrow p_3} + q^2 G_7
+
\mbox{perm.}
\end{align}

\begin{figure}[t]
\begin{center}
\mbox{
\begin{picture}(0,100)(60,0)
\put(0,0){\insertfig{4}{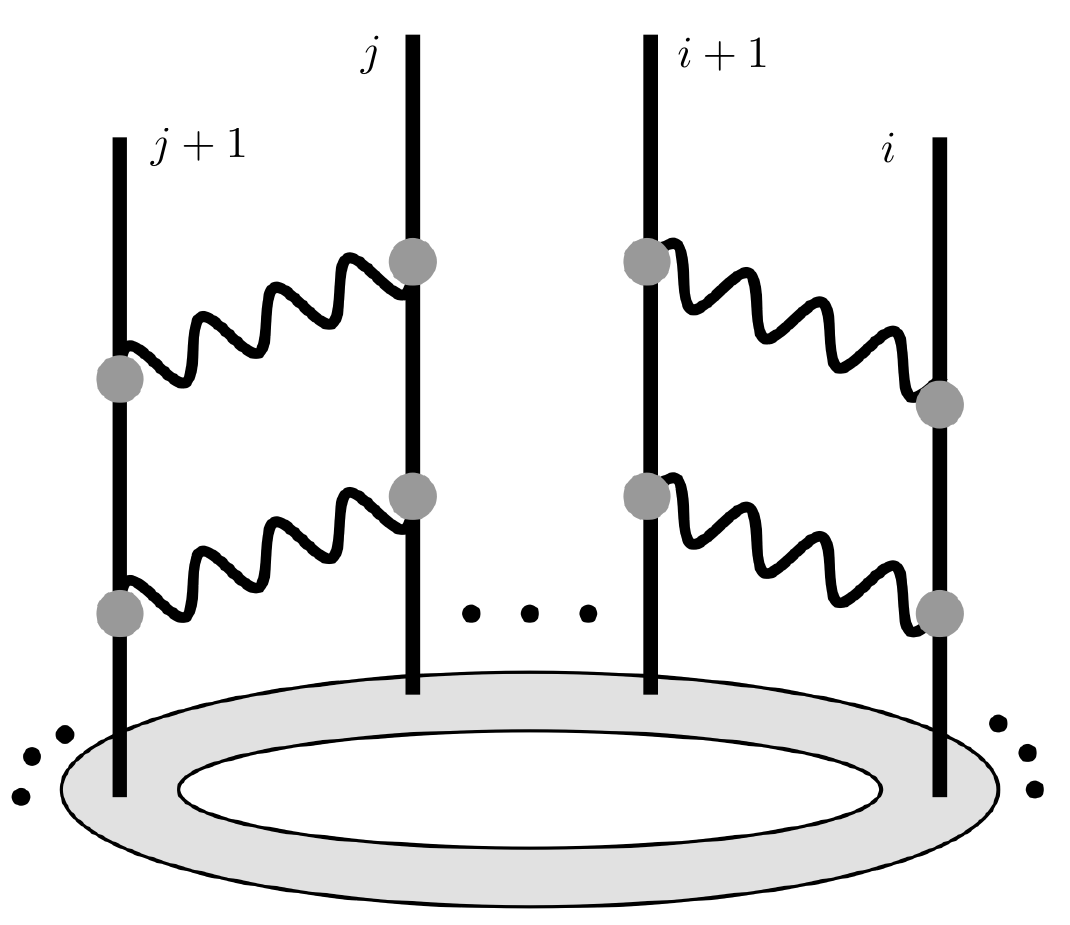}}
\end{picture}
}
\end{center}
\caption{\label{C1C1}  Factorised contribution of $\textbf{C}_1\times\textbf{C}_1$ supergraphs which will appear at two loop level for $n>3$ case.}
\end{figure}

Form factors for general $n>3$ can be considered along the same lines. The only new contribution comes from supergraphs 
where the two independent pairs of legs interact with each other, as shown in Fig.\ \ref{C1C1}. The final result in terms of scalar 
integrals is identical to (\ref{2loopFFNp}).

Concluding our considerations of the minimal off-shell form factors of the operators $\tr \Phi^n$ with $n$ external legs, we see that there 
is no difference between topologies of contributing scalar integrals as well as their accompanying coefficients compared to the on-shell 
case. Hence, this off-shell analysis agrees with the naive uplift to the Coulomb branch, at least in the $p_i^2 \to 0$ limit, at the two-loop 
level. It is an interesting question whether this equivalence persists at higher loops.

\section{Results}
\label{ResultsSection}

Having confirmed the set of Feynman integrals with two different methods, we are in a position to evaluate them analytically and
analyze their sum. From now on, the off-shellness is made dimensionless by dividing out $q^2$, i.e., $m/(-q^2)$. However,
to void writing this ratio in our subsequent formulas, we will simply set $q^2 = -1$.

\subsection{Solving integrals}

The one-loop form factor \re{1loopFF} is determined by the UD box function $\Phi_1 (x,y)$. Its near-mass-shell expansion 
in $m$, in the kinematics relevant to our discussion $x=y=m/z$ (with $z = \mbox{finite}$), is given by:
\begin{align}
\label{Box1loopExp}
\Phi_1\left(\frac{m}{z},\frac{m}{z}\right)
=
\log^2 m - 2\log m \log z + \log^2 z + 2 \zeta_2
+ 
O (m)
\, .
\end{align}
At two loops, as we emphasized earlier in Section \ref{UpliftSection}, the ${\rm DTri}_1$ and ${\rm DTri}_2$ integrals are equal to 
each other and can be expressed through the same two-loop UD box function $\Phi_2 (x,y)$. Its expansion for small $m$ reads
\begin{align}
\label{DBoxExp}
\Phi_2\left(\frac{m}{z},\frac{m}{z}\right) 
&= \ft14 \log^4 m - \log^3 m \log z 
+
\log^2 m \left( \ft{3}{2} \log^2 z + 3 \zeta_2 \right)
\\
&
-
\log m  \left( \log^3 z + 6 \zeta_2 \log z \right)
+
\ft{1}{4} \log^4 z + 3 \zeta_2 \log^2 z + \ft{21}{2} \zeta_4 
+ 
O (m)
\, . \nonumber
\end{align}

The $\mbox{TriBox}$ integral in Fig.\ \ref{fig2loop} (d) requires a dedicated analysis. Fortunately, it belongs to the family of double box integrals, 
which were recently studied in Ref.\ \cite{Belitsky:2023gba}. In that reference, a basis of 62 master integrals $J_i$ was constructed 
for the off-shell kinematics and they were solved in the near-mass-shell limit in terms of multiple polylogarithms \cite{Goncharov:2001iea}.
Therefore, all we need to do for our current consideration is to perform an integration-by-parts reduction of TriBox to $J_i$'s. This is easily
accomplished with the {\tt FIRE} package \cite{Smirnov:2023yhb} and the result is
\begin{align}
\label{TriBoxSolution}
{\rm TriBox}
=
\frac{u J_{31} + w (J_{14}-J_{27}+J_{28}-J_{36})
}{
\varepsilon^4 uw}
\, .
\end{align}
While the integral $\mbox{TriBox}_{\rm red}$ is simply given (up to an overall coefficient) by the 31-st element of the basis,
\begin{align}
\label{TriBoxRedSolution}
{\rm TriBox}_{\rm red}
=
\frac{J_{31}}{\varepsilon^4 w}
\, .
\end{align}
Notice the presence of the dimensional regularization parameter $\varepsilon$ in the denominator. The reason for this is that the 
basis of the 62 master integrals is built from elements some of which formally diverge in four dimensions. This choice is driven
by the desire to use the power of the canonical differential equations \cite{Henn:2013pwa} for systematic evaluation of Feynman 
integrals. Of course, their linear combinations in Eq.\ \re{TriBoxSolution} and \re{TriBoxRedSolution} are finite. This serves as a 
nice cross-check on the correctness of underlying manipulations. These results, expressed in terms of multiple polylogarithms are 
too lengthy to be displayed explicitly in the paper and are quoted in an accompanying {\tt Mathematica} notebook. Corresponding 
expressions were verified numerically using the {\tt GiNaC} integrator \cite{Bauer:2000cp} through the interactive {\tt Ginsh} 
environment of the {\tt PolyLogTools} package \cite{Duhr:2019tlz} against numerical evaluation of the parent momentum integrals 
using the {\tt FIESTA5} package \cite{Smirnov:2013eza,Smirnov:2021rhf}. It is also interesting to note that the individual results for 
${\rm TriBox}$ and ${\rm TriBox}_{\rm red}$ are not pure functions of Uniform Transcendentality (UT), i.e., the polylogs are
generally multiplied by rational functions of the Madelstam variables $u$, $v$ and $w$. Nevertheless, UT is restored in the
sum of all graphs as per anticipated properties of $\mathcal{N} = 4$ results. This is what we turn to in the next section.

\subsection{$n=3$ form factor}
\label{sec:part3}

According to the general conjecture \re{AmplFactorCB}, the three-leg minimal form factor admits the form
\begin{align}
\label{IRExpGen}
\log \mathcal{F}_3 
= 
- \frac{\Gamma_{\rm oct}(g)}{4}
\Big[\log^2\left(\frac{m}{u}\right) + \log^2\left(\frac{m}{v}\right)+ \log^2\left(\frac{m}{w} \right)\Big]
+
r_3 (u,v,w; g)+O(m)\, ,
\end{align}
with $r_3$ being the $m$-independent finite part. It will be convenient for our presentation, however, to subtract
out finite terms from the infrared logarithms in \re{IRExpGen} and instead decompose $\mathcal{F}_3$ as follows
\begin{align}
\label{IRExpGen1}
\log \mathcal{F}_3 
= 
- \frac{3}{4}\Gamma_{\rm oct}(g)\log^2 m + \frac{1}{2} \Gamma_{\rm oct}(g)\log m \log(uvw)
+
f_3 \left(u,v,w; g \right) + O(m)\, ,
\end{align}
with the remainder $f_3$ admitting the perturbative expansion
\begin{align}
\label{PTexpForf3leg}
f_3\left(u,v,w; g \right) = g^2 f_3^{(1)}(u,v,w) + g^4 f_3^{(2)}(u,v,w)+\ldots\, .
\end{align}

Now we are ready to test \re{IRExpGen1}. Substituting the near-mass-shell expansion for the DU integral \re{Box1loopExp}, we find 
that at one loop, the infrared logarithms do agree with Eq.\ \re{IRExpGen1} if we use the leading perturbative term in the octagon
anomalous dimension \re{ExactGammaOct}. While for the finite part, we get
\begin{align}
\label{f1}
f_3^{(1)}(u,v,w)=- \log^2 u - \log^2 v - \log^2 w + 6 \zeta_2
\, .
\end{align}

At two loops, we expect exponentiation of the infrared logarithms,
\begin{align}
\label{logF3exp}
\log \mathcal{F}_3 =g^2 \mathcal{F}_3^{(1)}+g^4 \left( \mathcal{F}_3^{(2)} - \ft{1}{2} \big[  \mathcal{F}_3^{(1)} \big]^2 \right)+\ldots.
\end{align}
The individual two-loop integrals in \re{2loopFF}, as can be seen from the attached {\tt Mathematica} notebook, are polynomials in 
powers of $\log^p m$ from $p=4$ downwards. However, we observe the cancellation of the top two powers of the infrared logs
in the two-loop combination \re{logF3exp}.

Now, we turn to the most tedious portion of the analysis, i.e., the two-loop finite part $f_{3}^{(2)}$. Unsimplified, the latter is given by 
a lengthy sum of multiple polylogarithms. As it became customary since the seminal work \cite{Goncharov:2010jf}, the use of the 
symbol map \cite{Goncharov:2009lql} provides a guiding `searchlight' on the path to a simplified representation. Applying this technology
to the current remainder function, we find that its symbol is given by the following one-line expression
\begin{align}
\mathcal{S}[ f_3^{(2)} ] =
u \otimes v \otimes  \left[ \frac{u}{w}\otimes \frac{v}{w} + \frac{v}{w}\otimes \frac{u}{w} \right] +
{1 \over 2}  u \otimes {u \over   (1-u)^3 }  \otimes \frac{v}{w}\otimes\frac{v}{w}   +  \text{perms}\, (u,v,w) \  ,
\end{align}
where ${\rm perms}\, (u,v,w)$ stands for {\sl all possible permutations} of $(u,v,w)$ (not only cyclic!). Such a symbol has already 
been encountered before in the computation of the finite remainder of the same form factor but for the purely massless case 
\cite{Brandhuber:2014ica}. A UT function of classical polylogarithms was lifted from it and found to be \cite{Brandhuber:2014ica}
\begin{align}
\label{Rem}
\mathcal{R}^{(2)}_3(u,v,w) 
= 
&
-
\frac{3}{2}\, \text{Li}_4(u)
+
\frac{3}{4}\,\text{Li}_4\left(-\frac{u v}{w}\right)
-
\frac{3}{2}\log w \, \text{Li}_3 \left(-\frac{u}{v} \right)
+ 
\frac{ 1}{16}  {\log}^2 u \log^2 v
\\
&+ {\rm perms}\, (u,v,w) \, .
\end{align}
So, all we are left to do is to find beyond-the-symbol contributions, if any. These terms are of the form zetas$\times$polylogs
and/or transcendental constants. The symbol is oblivious to them. A numerical evaluation of $f_3^{(2)}$ and $\mathcal{R}^{(2)}_3$ at
several kinematical points unambiguously indicates that $f_3^{(2)}$ and $\mathcal{R}^{(2)}_3$ are not the same and the difference is not 
merely a constant. This is not something unexpected. This happens in the conformal case \cite{Brandhuber:2014ica} as well as 
in the situation with the three particle form factor of the $\tr \phi_{12}^2$ operator, which we have previously considered \cite{Belitsky:2024agy}. 
To find the sought-after deviation from $\mathcal{R}^{(2)}_3$, we constructed an ansatz of all possible terms of the form $\{ \zeta_2 \log z_i \log z_j, 
\zeta_2 {\rm Li}_2 (z_i), \zeta_3 \log z_i, \zeta_4 \}$ accompanied by rational number coefficients. The $z_i$'s were taken from the 
alphabet\footnote{It would be much shorter if the cyclic symmetry as well as functional relations between ${\rm Li}_2$ were taken into account.}
\begin{align}
\left\{
u,v,w,1-u,1-v,1-w,1-\frac{1}{u},1-\frac{1}{v},1-\frac{1}{w}, -\frac{u v}{w}, -\frac{v w}{u},-\frac{w u}{v}
\right\} \, .
\end{align}
The unknown coefficients were then fixed by a high-precision $10^{-11}-10^{-13}$ numerical comparison of $f_3^{(2)} - \mathcal{R}^{(2)}_3$ 
against the ansatz using the {\tt Ginsh} integrator. The result we obtained is
\begin{align}
\label{f2}
f_3^{(2)} (u,v,w)
&
=
\mathcal{R}^{(2)}_3(u,v,w)
+
\ft{327}{8} \zeta_4 
+
\zeta_3 \left[\log u + \log v + \log w \right]
\\
&+
\ft{1}{4} \zeta_2 \left[ \log^2 u + \log^2 v + \log^2 w + 2 \log u \log v + 2 \log v \log w + 2 \log w \log u \right]
\, , \nonumber
\end{align}
and concludes our analysis of the off-shell three-leg minimal form factor.

Let us now discuss the key properties of the $n=3$ cas. Our explicit calculation at two loops confirms its exponentiation 
(\ref{AmplFactorCB}) adding another `notch to the belt' of the Coulomb branch. Its infrared behavior is governed by the octagon
anomalous dimension, not the cusp. At the level of the symbol, the conformal and Comlomb branches share the same
remainder function. They differ, however, at beyond-the-symbol contributions. Similar observations were made for the three-leg
form factor of the stress-tensor multiplet \cite{Belitsky:2024agy}. In that case, the difference was attributed to the inequivalent
collinear behavior of the two phases. Preliminary considerations suggest that the presence of the ultrasoft modes in loop
integrals yields discord with conventional wisdom. Minimal form factors do not possess conventional collinear properties already 
for the massless case since there is no lower-leg `observable' they can go to. So it is not a surprise that the finite remainder of 
Comlomb branch $n=3$ minimal form factor differs as well.

\subsection{$n>3$ form factors}

Finally, let us comment on the minimal form factors of the $\tr \phi_{12}^n$ operators for $n > 3$. In complete analogy to the $n=3$ case, 
the infrared logarithms exponentiate here as well. The finite part at one loop is a straightforward generalization of Eq.\ \re{f1} 
does not require any further consideration. As to the two-loop remainder $f^{(2)}_n$, we found that its symbol coincides with
its conformal counterpart \cite{Brandhuber:2014ica}
\begin{align}
\label{Symbfn}
\mathcal{S}[ f_n^{(2)}]
= 
\sum_{i=1}^n 
\bigg(
&
u_i \otimes (1-u_i)\otimes \Big[ \frac{u_i-1}{u_i} \otimes \frac{v_i}{w_i}
+ 
\frac{v_i}{w_i}\otimes \frac{w_i^2}{u_i v_i} \Big] 
\\ 
+ 
&
u_i\otimes u_i\otimes {1-u_i \over v_i} \otimes \frac{w_i}{v_i}  
+
u_i\otimes v_i\otimes  \Big[ \frac{v_i}{w_i} \otimes_S \frac{u_i}{w_i} \Big]
+ 
(u_i \leftrightarrow v_i ) \, \bigg) 
\, , \nonumber
\end{align}
with the variables $u_i,v_i,w_i$ defined by 
\begin{align}
u_i=\frac{s_{i\,i+1}}{s_{i\,i+1 \, i+2}} \, ,
v_i=\frac{s_{i+1\,i+2}}{s_{i\,i+1 \, i+2}} \, ,
w_i=\frac{s_{i+2\,i}}{s_{i\,i+1 \, i+2}} \, ,
\end{align}
and
\begin{align}
s_{i \, i+1 \, i+2} = s_{i \, i+1}+s_{i+1 \, i+2}+s_{i+2 \, i}
\, . 
\end{align}
They are generalisations of (\ref{mandelstam2}) and obey $u_i + v_i + w_i \simeq 1$. 

\section{Conclusions}

This work was dedicated to a two-loop analysis of off-shell minimal form factors in the $\mathcal{N} = 4$ sYM. We relied on
two approaches to determine the set of independent Feynman integrals that contribute to these `observables'. One relied
on the uplift of the conformal case to a higher dimension and interpretation of out-of-four-dimensional momenta as 
off-shellness $m$. The other was rooted in the $\mathcal{N} = 1$ superspace formulation of the model and off-shell analysis of 
contributing supergraphs. Both were shown to agree, in the near mass-shell limit, reinforcing our confidence in the correctness 
of our findings. We calculated all Feynman integrals as an expansion in the infrared logarithms of $m$ including the finite
remainder. 

We demonstrated the exponentiation of the Sudakov logarithms and confirmed the octagon anomalous dimension as a substitute 
for the conventional cusp anomalous dimension appearing in on-shell considerations. This makes it quite likely that the off-shell 
Sudakov behavior in QCD needs dedicated studies since off-shell partonic subprocesses enter as intrinsic building blocks of 
high-energy and $k_T$ factorization schemes.

The finite remainder was shown to be identical at the level of symbols to its conformal counterpart. It would be interesting to
generalize the current consideration to the three loops and see whether the symbols continue to agree at that order as well
\begin{align}
\label{RR}
\mathcal{S}[f_3^{(3)}(u,v,w)] \stackrel{?}{=} \mathcal{S}[\mathcal{R}_3^{(3)} (u,v,w)]
\, .
\end{align}
The latter was recently derived in Ref.\ \cite{Henn:2024pki}.

\begin{acknowledgments}
The work of A.B.\ was supported by the U.S.\ National Science Foundation under grant No.\ PHY-2207138. The work
of L.B.\ was supported by the Foundation for the Advancement of Theoretical Physics and Mathematics “BASIS”. 
\end{acknowledgments}

\providecommand{\href}[2]{#2}\begingroup\raggedright\endgroup

\end{document}